\documentclass[final]
  {aipproc}

\layoutstyle{6x9}

\newcommand{\beq}{\begin{equation}}
\newcommand{\eeq}{\end{equation}}
\newcommand{\bea}{\begin{eqnarray}}
\newcommand{\eea}{\end{eqnarray}}
\newcommand{\pt}{$p_T$}
\newcommand{\dd}{\mathrm{d}}

\begin{document}

\title{Dynamics and hadronization at intermediate transverse momentum at RHIC}


\author{V. Greco}{address={Dipartimento di Fisica e Astronomia, Via
    S. Sofia 64, I-95125 Catania, Italy}
}

\author{H. van Hees}{address={Cyclotron Institute and Physics Department, 
       Texas A\&M University, College Station, Texas 77843-3366, U.S.A.}
}

\author{R. Rapp}{address={Cyclotron Institute and Physics Department,
       Texas A\&M University, College Station, Texas 77843-3366, U.S.A.}
}

\begin{abstract}
  The ultra-relativistic heavy-ion program at RHIC has shown that at
  intermediate transverse momenta ($p_T \simeq 2$-$6$~GeV) standard
  (independent) parton fragmentation can neither describe the observed
  baryon-to-meson ratios nor the empirical scaling of the hadronic
  elliptic flow ($v_2$) according to the number of valence quarks. Both
  aspects find instead a natural explanation in a coalescence plus
  fragmentation approach to hadronization. After a brief review of the
  main results for light quarks, we focus on heavy quarks showing that a
  combined fragmentation and quark-coalescence framework is relevant
  also here. Moreover, within relativistic Langevin simulations we find
  evidence for the importance of heavy-light resonances in the
  Quark-Gluon Plasma (QGP) to explain the strong energy loss and
  collective flow of heavy-quark spectra as inferred from non-photonic
  electron observables. Such heavy-light resonances can pave the way to
  a unified understanding of the microscopic structure of the QGP and
  its subsequent hadronization by coalescence.
\end{abstract}

\maketitle

\section{Introduction}
Experimental data from the Relativistic Heavy Ion Collider (RHIC) during
the past few years have shown convincing evidence for a state of matter
with energy densities, $\epsilon$, in substantial excess of the expected
critical one, i.e., $\varepsilon \simeq 15$-$20 \varepsilon_c$ with
$\varepsilon_c\simeq 1$~GeV/fm$^{-3}$.  In the standard picture before
2003 the main concern in the study of produced hadrons at high
transverse momentum (\pt) was the underlying modification of partonic
spectra due to interactions in such a hot and dense medium, that would
reflect in a similar pattern for all hadrons through independent
fragmentation.  Indeed, one of the most exciting observations at RHIC
has been the suppression of high-\pt~particles in agreement with the
non-abelian radiative energy-loss theory within perturbative QCD
(pQCD)~\cite{Gyulassy:2004zy}.  However, for light hadrons the
observations of an anomalous baryon-to-meson production ratio at
intermediate \pt~up to $\simeq 6$~GeV and a scaling of the elliptic flow
with the number of quark constituents, has enforced revisions of an
independent fragmentation model for hadronization.  Instead coalescence
processes among massive quarks appear to be a convenient picture that
can naturally and quantitatively account for the main features of light
hadron production at intermediate
\pt~\cite{Greco:2003mm,Greco:2003vf,hwa,fries}.

For heavy quarks (charm ($c$) and bottom ($b$)), the energy loss
predicted by pQCD~\cite{armesto05} turned out to be insufficient (at
variance with the light quark case) to account for the observed large
nuclear suppression (small $R_{AA}$) and collectivity (large $v_2$) in
non-photonic single-electron
spectra~\cite{Kelly:2004,Adler:2005xv,bielcik06,adare07,abelev07}.  Here
the challenge is mainly in the understanding of the in-medium quark
interactions, even if the acquired knowledge on the hadronization
mechanism from light quarks plays a significant role as well.

Lattice QCD (lQCD) results suggest that resonance structures survive in
meson-correlation functions at moderate temperatures~\cite{asakawa04}
above $T_c$. Two of us have therefore suggested an effective model for
heavy-light quark scattering via $D$ and $B$
resonances~\cite{vanHees:2004gq}.  To study the consequences of such
interactions for the modification of the heavy-quark (HQ) distributions,
a Langevin simulation has been performed to trace the evolution of HQ
distributions through the fireball in heavy-ion Collisions
(HICs)~\cite{Teaney04,vanHees:2005wb}; hadronization has been modeled by a
coalescence model similar to the one applied to light
quarks~\cite{Greco:2003mm,Greco:2003vf}. It has been found that the
effect of resonant heavy-light scattering is
crucial~\cite{vanHees:2005wb} and also provides a reasonable agreement
with semileptonic $e^{\pm}$ spectra at RHIC
data~\cite{Adler:2005xv,abelev07,adare07}.

We note that for heavy quarks (heavy-light) quark-antiquark resonances
provide the dominant medium effects on their distributions close to
$T_c$, which then naturally merges into a coalescence-type description
for hadronization processes~\cite{Ravagli:2007xx}.

\section{Modification of hadronization mechanism}

\begin{figure}
\label{bm}
\includegraphics[height=.22\textheight]{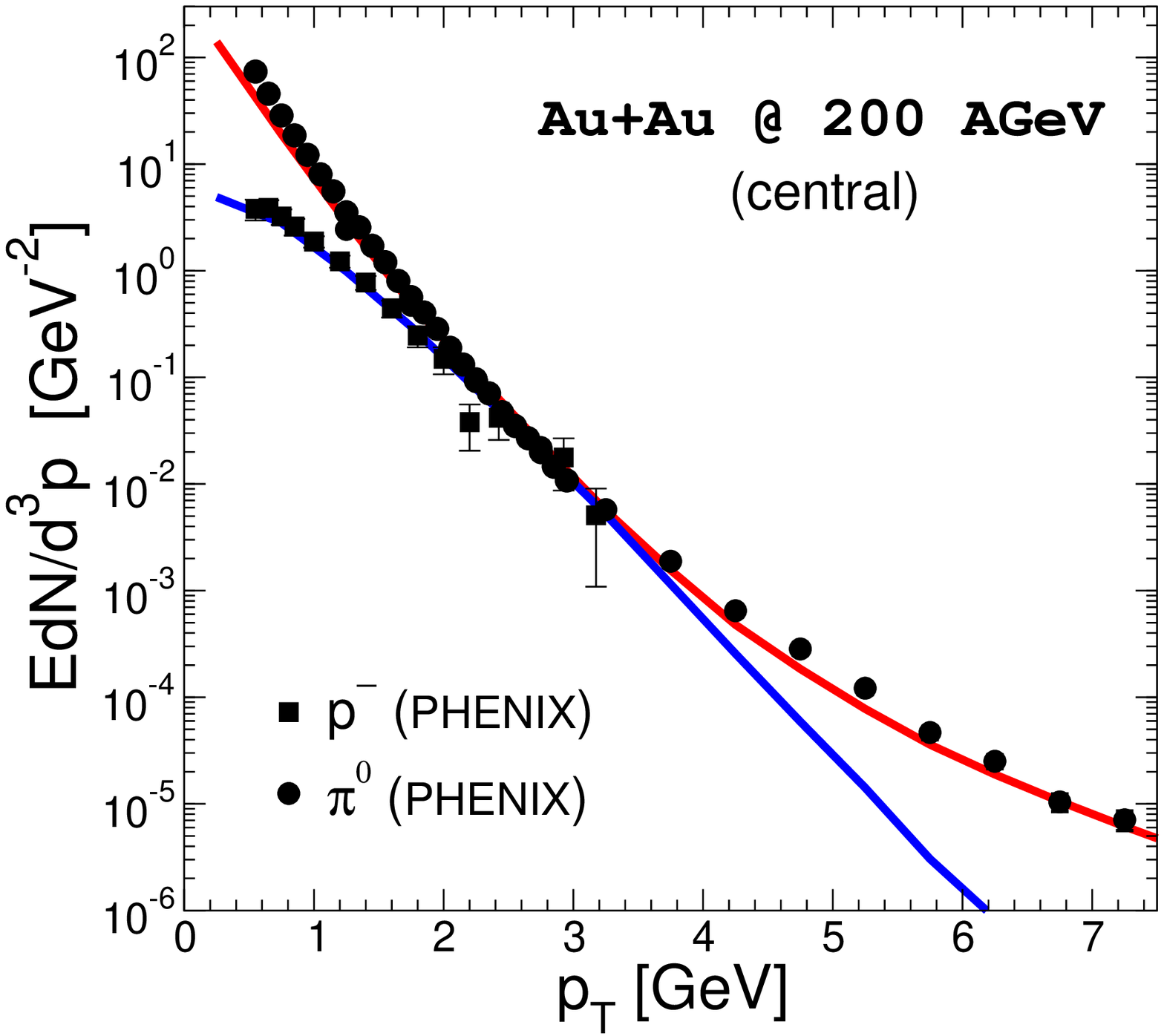}
\hspace*{0.3in}
\includegraphics[height=.22\textheight]{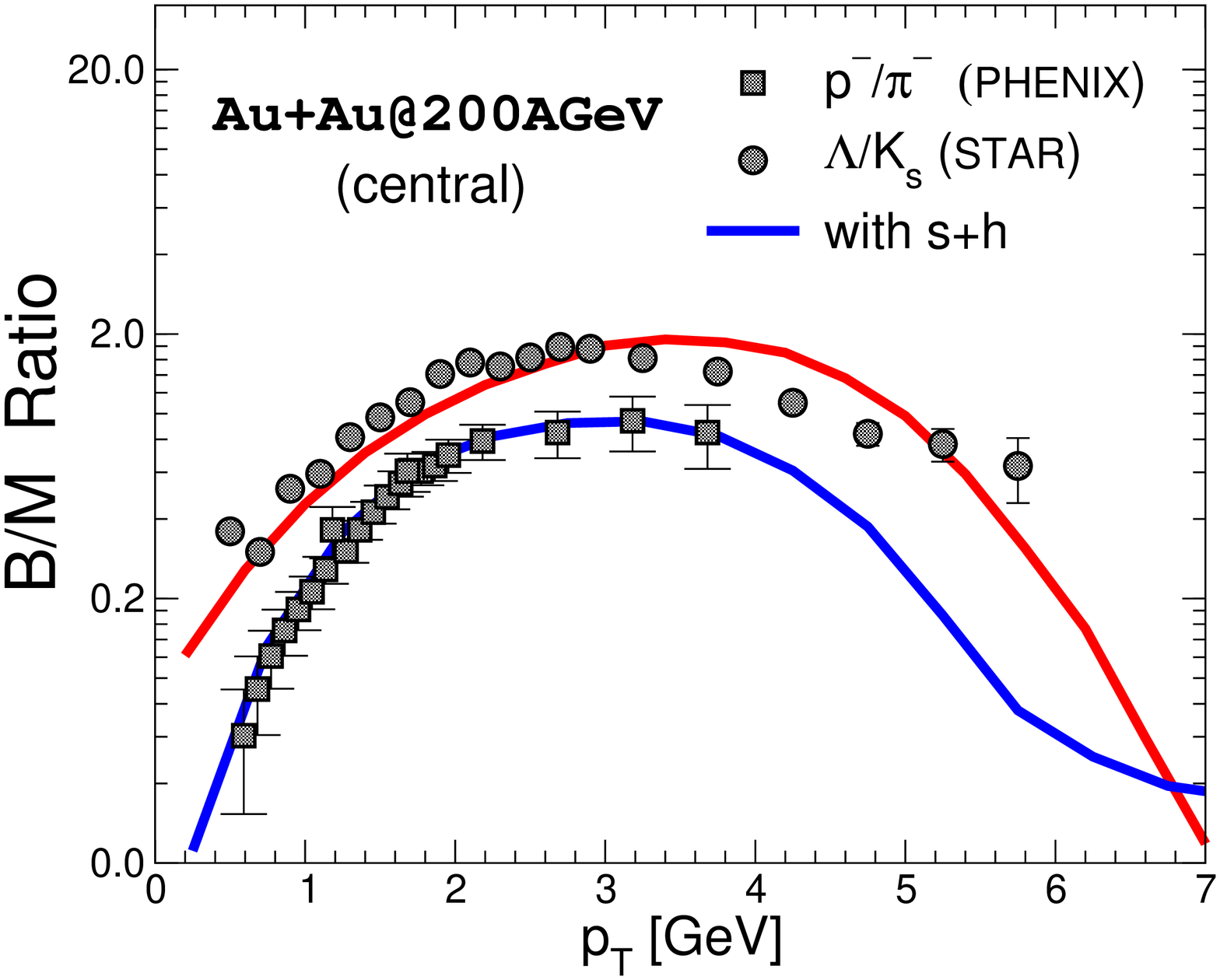}
\caption{Left: $\pi$ and $p^-$ transverse-momentum spectra in
  $\sqrt{s}$=200~AGeV Au+Au collisions. RHIC
  data~\cite{phenix-spec-high,phenix-spec-low} are shown by circles
  ($\pi^0$) and squares ($\bar{p}$), lines are results from coalescence.
  Right: experi\-ment vs. coalescence results for baryon-to-meson ratios
  for $p/\pi$ (lower part) and $\Lambda/K^0_{s}$ (upper part).}
\end{figure}

Hadronization at asymptotically large momentum can be described by a set
of fragmentation functions $D_{a/H}(p/P)$ that parametrize, in a
universal way, the probability that a hadron $H$ with momentum $P$ is
created from a parton $a$ with momentum $p$ in the vacuum.
Fragmentation functions have been measured in $e^{+}e^{-}$ collisions
and work well for hadron production at \pt > 2 GeV also in $pp$
collisions at RHIC energies.  Therefore it was expected that in this
\pt~regime the QGP could be probed by focusing on modifications of the
spectra, $E \dd N/\dd^3p$, at the parton level; but from Au+Au
collisions it became clear that this is not the case at least up to $p_T
\simeq 6$~GeV.  Two puzzling observations lead to this conclusion: (a)
baryons are much more abundant than predicted by fragmentation. A ratio
$\bar{p}/\pi \approx$ 1 between 2 and 4 GeV/c has been measured, see
Fig.~\ref{bm}, much larger than the value of $\approx$~0.2 predicted by
leading-twist pQCD. A similar trend is observed for $p/\pi$,
$\Lambda/K^0_{s}$, i.e., $p$, $\bar{p}$ and $\Lambda$'s do not seem to
suffer jet quenching. The pertinent nuclear modification factors,
$R_{AA}$, are close to 1, unlike those of light mesons for which $R_{AA}
\approx 0.2$ in central collisions; (b) the elliptic flow of all
identified hadrons is found to scale according to a quark-number scaling
as reflected in a universal behavior of $v_{2,H}(p_T/n)/n$ where $n$ is
the number of constituent quarks in hadron $H$.  In particular, recent
data for the $\phi(1020)$ also follow the scaling, suggesting the
dominance of the quark content rather than the mass effect, in agreement
with the coalescence prescription~\cite{Afanasiev:2007tv}.

The main reason for the inadequacy of a pure fragmentation picture is
the high density of the matter created in HICs. In such an environment
one may expect that quarks could just coalesce into hadrons: three
quarks into a baryon, a quark-antiquark pair into a meson.  In such a
picture baryons with momentum \pt~are mainly produced from quarks with
momenta $\simeq$\pt/3, while mesons with the same momentum mainly arise
from quarks with momenta $\simeq$\pt/2. This is contrary to the
fragmentation process where baryon production is suppressed with respect
to mesons as more quarks are needed from the vacuum. A coalescence model
that is based on the simple overlap of the quark-distribution function
with a hadron-wave function has been developed to implement the physical
ideas sketched above~\cite{Greco:2003mm}.  In such a model the
transverse-momentum spectrum of hadrons that consists of $n$ (anti-)
quarks is given by the overlap between the hadron-wave function and $n$
quark phase-space distribution functions, $f_{q}(x_{i},p_{i})$:
\begin{equation}
\frac{\dd N_{H}}{\dd^2P_T}= g_H \int \prod_{i=1}^{n}\frac{\dd^{3}\mathbf{p}_{i}}
{(2\pi)^{3}E_{i}}{p_{i}\cdot 
\dd\sigma _{i}} f_{q}(x_{i},p_{i})f_{H}(x_{1}..x_{n};p_{1}..p_{n})\,\delta^{(2)}
\left(P_T - \sum_{i=1}^n p_{T,i}\right)   \  ;
\label{coal1}
\end{equation} 
$\dd\sigma$ denotes an element of a space-like hadronization
hypersurface, $f_H$ is the Wigner distribution function of the hadron
and $g_H$ the probability of forming a color-neutral object with the
spin of the hadron from $n$ colored quarks. Therefore it is assumed that
the probability of coalescence is simply given by the phase-space
distance weighted by the wave function of the produced particle
Eq.~(\ref{coal1}).  In such an approach constituent-quark masses are
included representative for non-perturbative effects. This is a further
assumption that can be relaxed for heavy quarks owing to the smaller
effect of the QCD vacuum on their masses.  The distributions
$f_q(x_i,q_i)$ are fixed as homogeneous Boltzmann distributions with an
average radial flow $\langle\beta \rangle =$0.35 for $p_T < 2$ GeV, and
quenched minijets for $p_T > 2$ GeV. The volume is fixed to reproduce
the measured transverse energy at given centrality.  In the left panel
of Fig.~\ref{bm}, the $p$ and $\pi$ spectra obtained from the
quark-coalescence model are shown together with the experimental data
from PHENIX~\cite{phenix-spec-high,phenix-spec-low}.

The resulting $p/\pi$ ratio is shown in the right panel of Fig.~\ref{bm}
by lower lines \cite{ppi}, a similar effect is seen also for the
$\Lambda/K^0_{s}$ ratio (upper lines) \cite{lamont04}. Similar
conclusions are reached in other studies based on quark coalescence
\cite{hwa,fries}.

\section{Heavy-quarks at high temperatures}
Heavy quarks ($b,c$) are produced out of thermal equilibrium in the very
early stage of the reaction; due to their large mass a perturbative
evaluation of their in-medium interactions was expected to be reliable
also at relatively small \pt. However, a small nuclear modification
factor, $R_{AA} \simeq 0.3$, has been deduced from semileptonic electron
spectra associated with decays of $D$- and
$B$-mesons~\cite{Adler:2005xv,abelev07}, comparable to the pion one.
Such a value is incompatible with pQCD jet-quenching
mechanisms~\cite{armesto05}. This statement is strengthened by the
observed $v_2$ of up to 10$\%$~\cite{Kelly:2004,adare07}, indicating
substantial collective behavior of charm ($c$)
quarks~\cite{Greco:2003vf}. Moreover, a consistent description of
$R_{AA}$ and $v_2$ cannot be achieved even if one artificially upscales
the transport coefficients within pQCD energy-loss
calculations~\cite{Teaney04}. This suggests that the physics underlying
the heavy quark observables is not only a matter of a global evaluation
of the interaction strength, but there is an opportunity for a more
detailed understanding of the microscopic nature of the interaction
(most likely of non-perturbative origin) and of the hadronization
mechanism, as we briefly review in the following.

A hint on non-perturbative interactions of heavy quarks in the medium is
provided by lQCD computations which exhibit resonance structures in
meson correlation functions at moderate
temperatures~\cite{asakawa04,KL03}.  Along this line two of us have
suggested that $D$- and $B$-resonance exchange in the $\bar q -Q $
channel may be the dominant scattering process that drives the HQ
dynamics~\cite{vanHees:2004gq}.  To evaluate the consequences of such a
picture we have built a model based on an effective Lagrangian:
\begin{equation} 
{\cal L} = Q~\frac{1+\not{\!v}}{2}~\Phi~\Gamma~\bar q +
  \mathrm{h.c.}
\label{lag}
\end{equation}
where $\Phi$=$D$, $B$. We have calculated elastic $Q+\bar q \to Q+\bar
q$ scattering amplitudes via $\Phi$ exchange in the $s$- and
$u$-channel.  The existence of one $\Phi$ state (e.g., a pseudoscalar
$J^P$=$0^-$), is assumed together with a minimal degeneracy following
from chiral and HQ symmetries, represented by Dirac matrices $\Gamma$=1,
$\gamma_5$, $\gamma^\mu$, $\gamma_5\gamma^\mu$ in Eq.~(\ref{lag}).

The application to HICs is realized by treating HQ kinetics in the QGP
as a relativistic Langevin process~\cite{vanHees:2005wb}:
\begin{equation}
  \frac{\partial f}{\partial t} = \frac{\partial (\gamma p f)}{\partial p}
  + \frac{\partial^2 (D_p f)}{\partial p^2} \ ; 
\end{equation}
$\gamma$ and $D_p$ are drag and (momentum) diffusion coefficients which
determine the approach to equilibrium and satisfy the Einstein relation,
$T=D_p/\gamma M_Q$.  The bulk medium is modeled by a spatially
homogeneous elliptic thermal fireball which expands isentropically.
Finally, hadronization is treated via coalescence at $T_c$=180~MeV, see
Eq.~(\ref{coal1}), plus fragmentation processes evaluated as
$f_{c,b}(p_T)*[1-P_{c,b\rightarrow(D,\Lambda_c),(B,\Lambda_b)}(p_T)]$,
where $P_{c,b\rightarrow (D,\Lambda_c),(B,\Lambda_b)}$ is the
probability for a heavy quark to coalesce.

\begin{figure}
\includegraphics[height=.21\textheight]{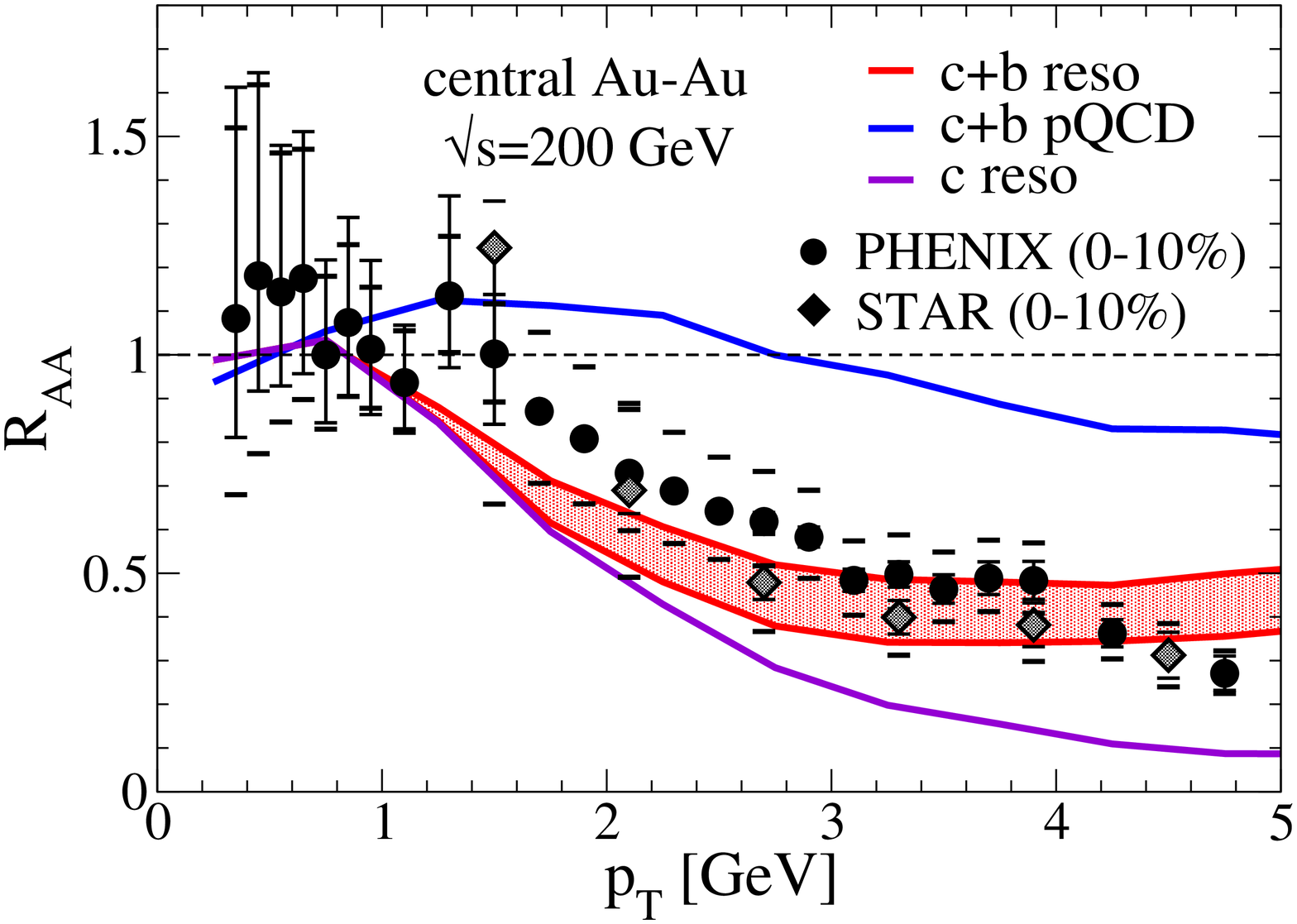} \hspace*{5mm}
\includegraphics[height=.21\textheight]{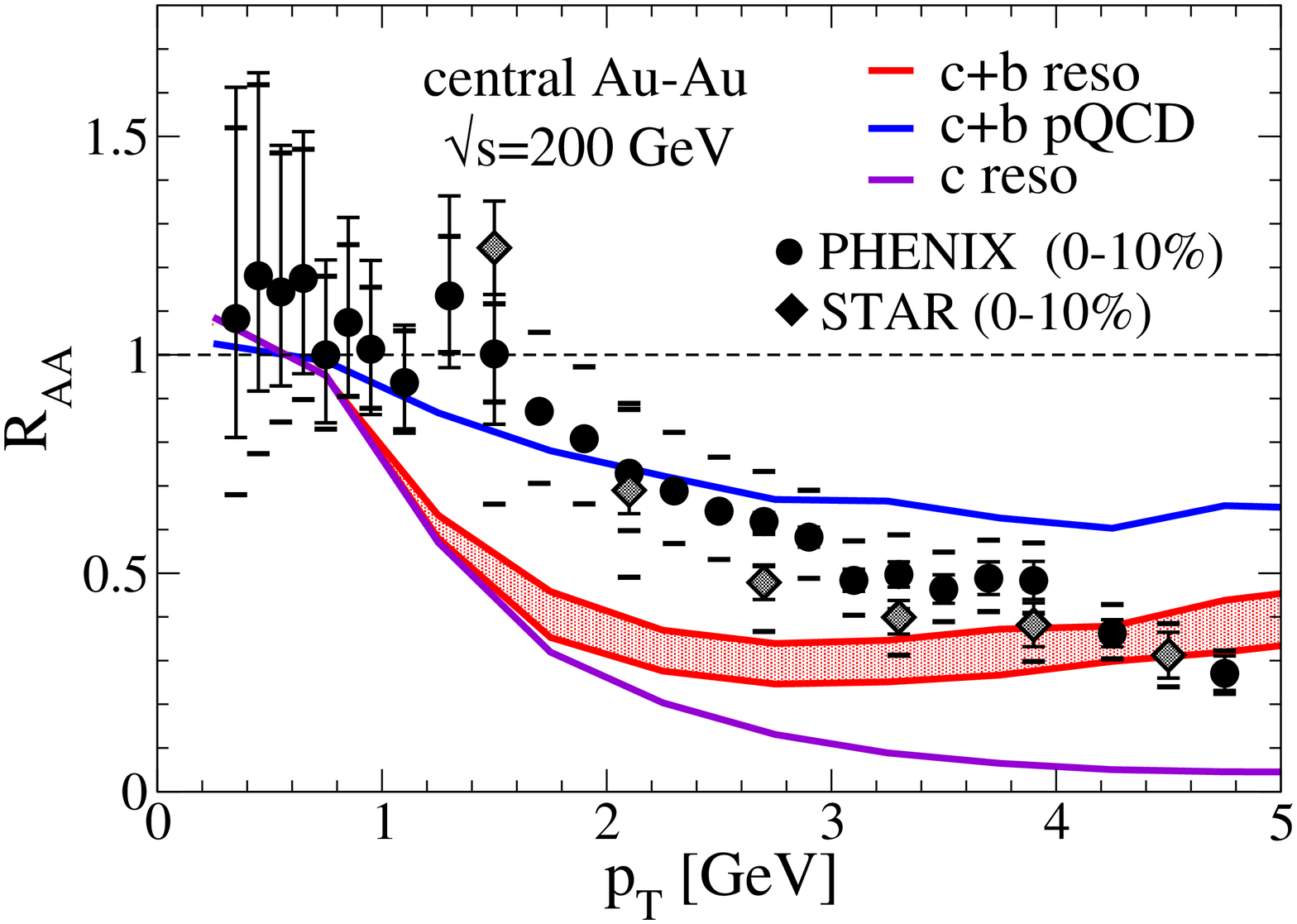}
\caption{Nuclear modification factor for single electrons, including
  coalescence and fragmentation at hadronization (left panel) and only
  with fragmentation (right panel), compared to RHIC 
  data~\cite{adare07,abelev07}.}
\label{raa}
\end{figure}

For 200~AGeV Au+Au collisions results from the Langevin simulation
including hadronization by coalescence+fragmentation (left) and
fragmentation only (right) are shown in Fig.~\ref{raa} together with
experimental data~\cite{Adler:2005xv,bielcik06}. It is obvious that
elastic scattering in a pQCD scheme is insufficient to account for the
small $R_{AA}$, independent of the hadronization scheme applied.  The
red band shows the full calculation with $c,b$ quarks that scatter in
the presence of hadron-like resonances with a width $\Gamma \simeq
0.4$-$0.75$~GeV (representing the interaction strength), supplemented by
the pQCD elastic scattering in color non-singlet channels (dominated by
gluons).  We note that the contamination of single electrons from $B$
decays is significant already at $p_T \sim 2$ GeV (corresponding to a
crossing of $c$ and $b$ spectra at around \pt$\sim$4-5~GeV). Thus the
inclusion of $B$ mesons (despite the inherent uncertainties in the
$b$/$c$ ratio) is mandatory to draw reliable conclusions on the
interaction processes underlying the experimental results.

\begin{figure}
 \includegraphics[height=.21\textheight]{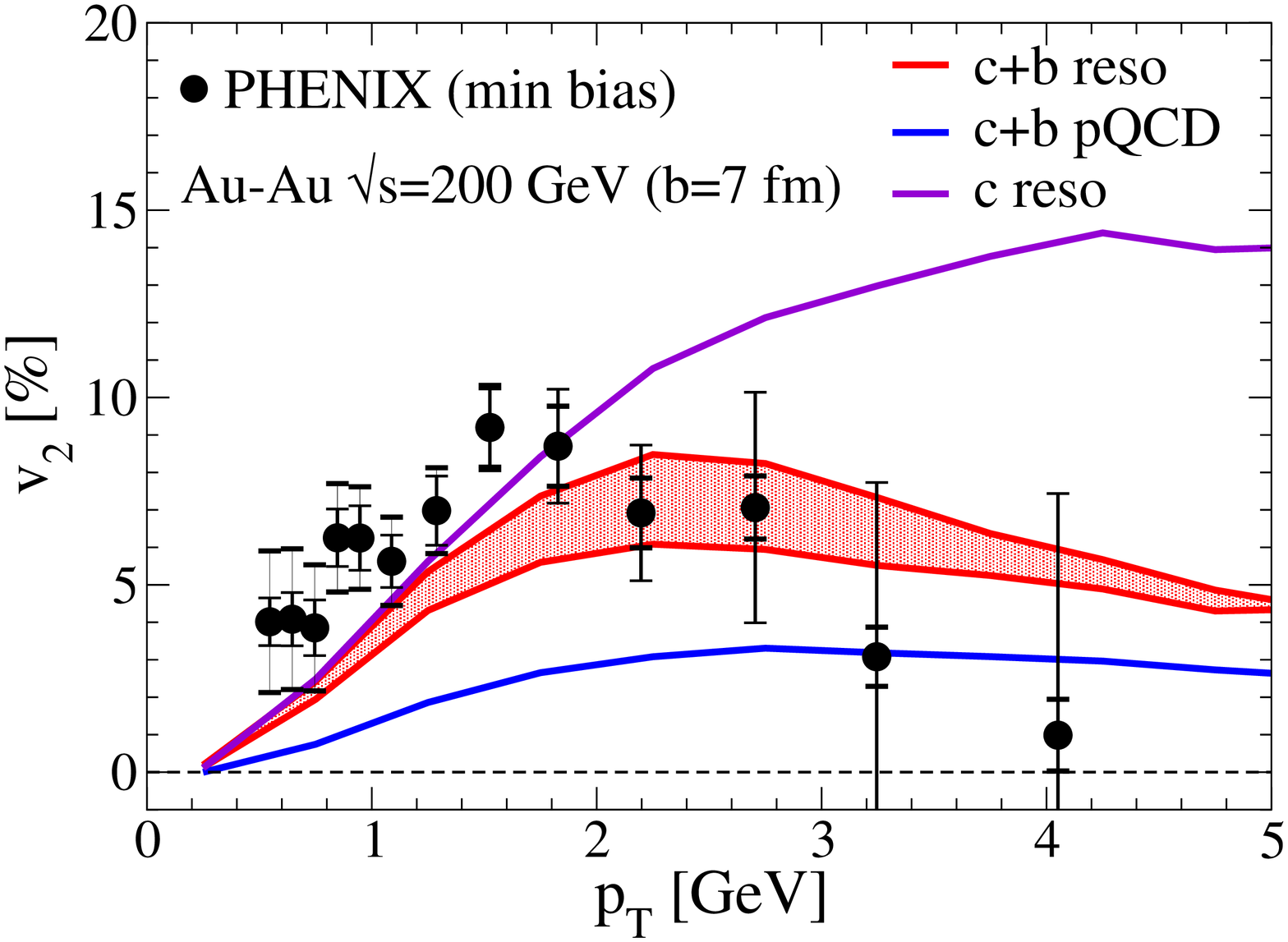}
 \hspace*{5mm}
 \includegraphics[height=.21\textheight]{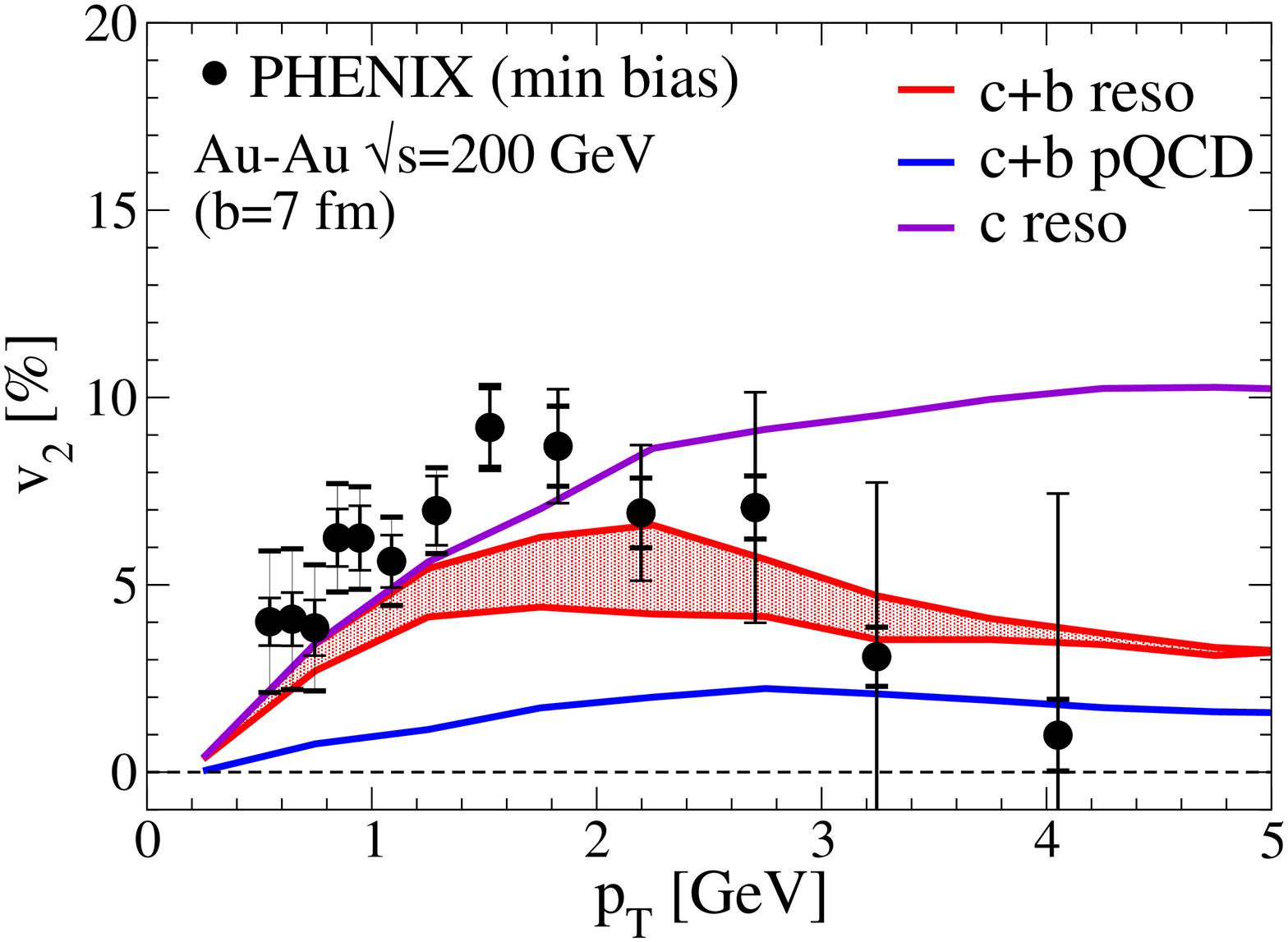}
 \caption{Elliptic flow for single electrons, including coalescence and
   fragmentation at hadronization (left panel) and only with
   fragmentation (right panel), compared to RHIC data~\cite{adare07}.}
\label{figv2}
\end{figure}

When comparing the band in the left and right panel of Fig.~\ref{raa} we
can see a clear coalescence effect of hadronization in terms of an
increase in the $R_{AA}$ at $p_T \sim 1$-$4$ GeV. This effect becomes
more significant via a simultaneous enhancement of the elliptic flow
$v_2$ \cite{Greco:2003vf,vanHees:2005wb}, see Fig. \ref{figv2} This behavior is typical of a
coalescence mechanism, reversing the usual correlation between $R_{AA}$
and $v_2$, and allows for a reasonable description of the experimental
data on both $R_{AA}$ and $v_2$.
Very recently, a Brueckner many-body scheme for in-medium T-matrices
for HQ scattering off light quarks \cite{van Hees:2007me} has
been evaluated starting from lQCD potentials.  The existence of $D$ and
$B$-like resonances has been confirmed; furthermore, when embedding 
the pertinent drag and diffusion coefficients in the model described 
above, a comparable (or even better) agreement with the data 
is found~\cite{van Hees:2007me}.

\section{Conclusions}
The first stage of RHIC program has shown clear signs of modifications
of the hadronization mechanism in the light-quark sector relative to
$pp$ collisions. There are several evidences that hadronization proceeds
through the coalescence of (massive) anti-/quarks which are close in
phase space.  The modification of hadronization for light quarks also
seems to play a role in the new challenge posed by heavy-quark probes.
Here, the main issue is the dominant interaction mechanism (HQ
diffusion) and its relation to the microscopic structure of the QGP. We
have presented a scenario based on the existence of $B$- and $D$-like
resonances in the QGP up to $T = 2\,T_c$.  Reasonable agreement with
experimental data is achieved owing to a positive interplay between the
resonance scattering mechanism and the hadronization by coalescence.  We
note that such an approach provides an inherent consistency between the
in-medium interactions of HQs and the subsequent hadronization by
coalescence: a pole in the quark-antiquark propagator above $T_c$ can be
viewed as a precursor of recombination.

Finally, it is important to keep in mind the interrelations of open and
hidden charm, if regeneration makes up a good fraction of the charmonium 
yield as expected at RHIC and LHC. In this case, the study of $B$ and $D$ 
distributions can be related
to that of $J/\psi$ and $\Upsilon$ due to the underlying common $c$, $b$
distributions; the consistency of such a picture should be checked in
the near future.

\begin{theacknowledgments}
Work of HvH + RR supported by 
the U.S. NSF under under contract no. PHY-0449489.
\end{theacknowledgments}

\bibliographystyle{aipproc}   

\end{document}